\documentclass[11pt,a4paper]{article}
\pdfoutput=1
\usepackage{jheppub}

\usepackage{amsmath}
\usepackage{verbatim}
\usepackage{amssymb}
\usepackage{inputenc, array}
\usepackage{textcomp}
\usepackage{bm}
\usepackage{physics}
\usepackage{simpler-wick}
\usepackage{appendix}
\usepackage{xcolor}

\newcommand{\e}{\epsilon}
\newcommand{\be}[1]{\begin{equation}\label{#1} }
\newcommand{\ee}{\end{equation}}
\newcommand{\bea}[1]{\begin{eqnarray}\label{#1} }
\newcommand{\eea}{\end{eqnarray}}

\renewcommand{\O}{{\mathcal{O}}}
\renewcommand{\L}{{\mathcal{L}}}

\newcommand{\h}{{\bar h}}

\renewcommand{\>}{\rangle}
\newcommand{\<}{\langle}
\newcommand{\w}{\omega}

\renewcommand{\th}{\theta}

\newcommand{\D}{\Delta}

\renewcommand{\t}{\tau}
\newcommand{\s}{\sigma}

\newcommand{\y}{\textbf{y}}
\usepackage{array}
\usepackage{multirow}
\usepackage{amsmath}
\usepackage{makecell}

\definecolor{green}{rgb}{0.1,0.8,0.2}

\usepackage{cancel}
\usepackage{graphicx} 

\usepackage{hyperref}
\usepackage{cleveref}
\usepackage{float}
\usepackage{subcaption}
\usepackage{physics}
\usepackage{slashed}
\usepackage{amsfonts}
\usepackage{caption}
\usepackage{mathrsfs}

\usepackage{jheppub}
\usepackage{tikz}
\usepackage{dirtytalk}
\usetikzlibrary{decorations.markings}
\usepackage{amsmath}
\usepackage{amssymb}
\usepackage{comment}
\usepackage{physics}
\usepackage{mathrsfs}
\usepackage{soul}

\title{BMS modular covariance and  structure constants}
\author[a, b, c]{Arjun Bagchi,} \author[a]{Saikat Mondal,} \author[a]{Sanchari Pal} \author[d]{and Max Riegler} \author{\\}

\affiliation[a]{Indian Institute of Technology Kanpur, Kalyanpur, Kanpur 208016. India.\\}
\affiliation[b]{Centre de Physique Théorique, Ecole Polytechnique de Paris, 91128 Palaiseau Cedex, France.\\}
\affiliation[c]{NORDITA,  Hannes Alfv{\'e}ns v{\"a}g 12, 106 91 Stockholm, Sweden\\}
\emailAdd{abagchi@iitk.ac.in, saikatmd@iitk.ac.in, pals@iitk.ac.in, max.riegler@tuwien.ac.at}
\affiliation[d]{Institute for Theoretical Physics, TU Wien,Wiedner Hauptstrasse 8-10/136, 1040 Vienna, Austria.\\}

\abstract{Two-dimensional (2d) field theories invariant under the Bondi-Metzner-Sachs algebra, or 2d BMSFTs in short, are putative holographic duals of Einstein gravity in 3d asymptotically flat spacetimes. When defined on a torus, these field theories come equipped with a modified modular structure. We use the modular covariance of the BMS torus two-point function to develop formulae for different three-point structure constants of the field theory. These structure constants indicate that BMSFTs follow the eigenstate thermalization hypothesis, albeit with some interesting changes to usual 2d CFTs. The singularity structures of the structure constants contain information on perturbations of cosmological horizons in 3d asymptotically flat spacetimes, which we show can also be obtained as a limit of BTZ quasinormal modes.}

\begin{document}
\maketitle

\section{Introduction} 
The success of the AdS/CFT correspondence \cite{Maldacena:1998zhr,Witten:1998qj,Gubser:1998bc} leading to a deeper understanding of theories of quantum gravity in terms of lower-dimensional quantum field theories calls for a more generalized understanding of the holographic principle. In this context, understanding holography for asymptotically flat spacetimes is particularly important since most of the real-world physical processes, e.g., processes involving astrophysical black holes, happen in approximately flat spacetime. 

\medskip

One of the precursors of the AdS/CFT correspondence was the analysis of asymptotic symmetries of AdS$_3$ in Einstein gravity by Brown and Henneaux \cite{Brown:1986nw}. They showed that with suitable boundary conditions, the symmetries at the asymptotic boundary are enhanced to two copies of the infinite-dimensional Virasoro algebra, thereby providing a connection between gravity and two-dimensional conformal field theories. These infinite-dimensional asymptotic symmetry algebras are not unique to AdS$_3$ but were also found in asymptotically Minkowski spacetimes in the 1960's by Bondi, van der Burg, Metzner \cite{Bondi:1962px} and independently by Sachs \cite{Sachs:1962zza}. They showed that in asymptotically flat four-dimensional (4d) spacetime, the symmetry algebra at null infinity gets enhanced from the usual Poincare algebra to an infinite dimensional algebra now called $\mathrm{BMS}_{4}$ algebra and is given by 
\begin{align}
\label{BMS4}
    [L_{n},L_{m}]&=(n-m)L_{m+n},\quad [\Bar{L}_{n},\Bar{L}_{m}]=(m-n)\Bar{L}_{m+n},\nonumber\\
    [L_{m},M_{p,q}]&=\left(\frac{m+1}{2}-p\right)M_{m+p,q},\quad
    [\Bar{L}_{m},M_{p,q}]=\left(\frac{m+1}{2}-q\right)M_{p,m+q},\nonumber\\
    [M_{p,q},M_{r,s}]&=0,
\end{align}
where $n,m=0,\pm 1$ and all other indices run through all integers. $M_{r,s}$'s are the generators of supertranslations that depend on the angles of the celestial sphere at null infinity and form an infinite-dimensional subgroup. The $L_{n}$'s are the Lorentz generators. The sub-algebra spanned by $L_{n}$ and $\Bar{L}_{n}$ gets enhanced to two copies of the Witt algebra if one relaxes the condition that generators need to be globally well defined \cite{Barnich:2010eb}. This extension of the BMS$_4$ algebra has been used predominantly in the literature when considering aspects of holography in 4d asymptotically flat spacetimes. {\footnote{Other extensions, including the enhancement of the global conformal group on the sphere to the full diffeomorphism group Diff(S$^2$) \cite{Campiglia:2014yka} have also been investigated.}} 

\medskip

The investigation of holography in asymptotically flat spacetimes has seen a resurgence in recent years and efforts have followed two distinct paths now called Celestial and Carrollian holography. Celestial holography \cite{Strominger:2013jfa, He:2014laa, Strominger:2014pwa} makes use of the fact that the Lorentz group in the bulk of flat space acts as the global conformal group on the null boundary and has been very successful in a novel understanding of S-matrices in terms of a dual 2d Celestial CFT that lives on the celestial sphere \cite{Pasterski:2016qvg,Pasterski:2017kqt,Banerjee:2018gce}. This approach yielded lots of new physics about scattering amplitudes and asymptotic symmetries. We refer the reader to the wonderful reviews \cite{Raclariu:2021zjz,Pasterski:2021rjz,Strominger:2017zoo} for more details. 

\medskip

Carrollian holography, on the other hand, seems to be more natural from the point of view of the original ideas of holography and proposes a co-dimension one holographic dual where one not only uses the Lorentz sub-group of the Poincare group but fits translations naturally into the formulation \cite{Bagchi:2016bcd}. Recent successes include \cite{Donnay:2022wvx,Bagchi:2022emh,Donnay:2022aba,Bagchi:2023fbj,Nguyen:2023vfz,Saha:2023hsl}. The initial success of the formulation was in lower dimensions, specifically in understanding 3d asymptotically flat spacetime, which we describe below. 

\medskip 

In three dimensions, the asymptotic symmetry algebra is infinite-dimensional and is given by the centrally extended BMS$_{3}$ algebra \cite{Barnich:2006av}:
\begin{align}
\label{cca}
   [L_{n},L_{m}]&=(n-m)L_{n+m}+c_{L}(n^{3}-n)\delta_{n+m,0},\nonumber\\
   [L_{n},M_{m}]&=(n-m)M_{n+m}+c_{M}(n^{3}-n)\delta_{n+m,0},\nonumber\\
   [M_{n},M_{m}]&=0. 
\end{align}
In the above, $M_n$'s are the generators of supertranslations that depend on one index corresponding to the Fourier modes of the celestial circle at null infinity (as opposed to the 2-sphere in the 4d case). The $L_n$'s are the superrotations that form the algebra of diffeomorphisms of the celestial circle at null infinity. This algebra admits central extensions $c_L$ and $c_M$. For Einstein gravity, $c_L=0$ and $c_M=3/G$ \cite{Barnich:2006av}. One can have non-zero values of $c_L$ when considering theories beyond Einstein gravity, e.g., Topologically Massive Gravity \cite{Bagchi:2012yk}. 

\medskip

Motivated by the construction of Brown and Henneaux and its subsequent understanding in terms of AdS$_3$/CFT$_2$, the basic observation of flat space holography in this formulation is to propose that the algebra \eqref{cca} also governs the putative 2d dual field theory living on the null boundary of flat spacetime \cite{Bagchi:2010zz,Bagchi:2012cy}. This algebra can also be obtained by taking the speed of light to zero on the dual CFT \cite{Bagchi:2012cy}, which motivated the recent nomenclature of Carrollian holography, following the Carrollian contraction of the Poincare algebra first obtained in the 1970s \cite{galileigroup,Levy}. The process of getting from AdS to flat space by sending the radius of AdS to infinity corresponds to sending the speed of light to zero in the dual field theory. 

\medskip

This correspondence between 3d bulk asymptotically flat spacetimes and 2d BMS$_3$ invariant field theories or 2d Carrollian CFTs has been quite successful during the past decade, with several checks of quantities computed between the bulk and boundary theories \cite{Bagchi:2012xr,Barnich:2012xq,Bagchi:2014iea,Barnich:2015mui,Bagchi:2015wna,Jiang:2017ecm,Hijano:2017eii}. One of the main avenues of success has been the consideration of modular properties of these Carroll CFTs and consequences for the analogs of the BTZ black holes \cite{Banados:1992wn,Banados:1992gq} in 3d flat space which turn out to be cosmological solutions called Flat Space Cosmologies (FSCs). Modular invariance of the Carroll partition function leads to the BMS-Cardy formula and a subsequent matching of the entropy of FSCs counting microstates of the BMSFT on the boundary  \cite{Bagchi:2012xr,Bagchi:2013qva,Bagchi:2019unf}. Modular covariance of the torus one-point function has resulted in the understanding of the structure constants, which were also reproduced by a bulk calculation involving a probe in the FSC geometry \cite{Bagchi:2020rwb}.

\medskip

In this paper, we extend our explorations of modular aspects of 2d Carrollian CFTs to the study of torus two-point functions. Our interest in this study is two-pronged. The first concerns an exploration of the thermal properties of Carrollian CFTs. In a 2d CFT, the torus two-point function naturally leads to the off-diagonal structure constants of the theory \cite{Brehm:2018ipf, Hikida:2018khg,Romero-Bermudez:2018dim}, which in turn provides an interpretation in terms of the Eigenstate Thermalization Hypothesis (ETH) \cite{Srednicki:1995pt}. ETH is a criterion that determines whether a closed quantum system thermalizes at late times under unitary evolution. The study of Carrollian field theories has and continues to reveal many aspects of these theories that are unfamiliar from the perspective of usual relativistic QFTs. Among them is the question of thermal equilibrium. Our study of the torus two-point functions in this work provides some hints of a notion of ETH emerging in 2d Carroll CFTs, albeit with some interesting differences.    



\medskip

Our final interest in this paper is understanding the singularity structure of the BMS torus two-point functions that we compute. For the 2d CFT case, the singularities of the torus two-point function carry information on the quasi-normal modes of the dual BTZ black hole. In the Carrollian case, which we primarily focus on in the paper, the structure of poles should shed light on the quasi-normal modes of FSC solutions in 3d flat spacetimes. We find interesting singularity structures from different types of torus two-point functions and comment on their relationship to physics in the bulk. 

\medskip

The paper is organized as follows. We begin in Sec.~2 with a compact review of 3d flat holography from the Carrollian perspective and elaborate modular aspects of the 2d BMSFTs. Sec.~3 contains the main computations of this work. Here we derive the off-diagonal structure constants by using modular covariance of the BMS torus two-point functions. We arrive at two distinctive formulae. One for temporally separated probes leading to a type of averaged three-point coefficient, while the other for spatially separated probes leads to a different class off-diagonal structure constant. In Sec.~4 and 5, we dwell on the physics arising from these two different types of structure constants.  In Sec.~4, we investigate the Eigenstate Thermalization Hypothesis, and after a brief detour to 2d CFTs, we show that ETH holds for 2d BMSFTs as well, with some interesting changes. In Sec.~5, we investigate the analogs of quasinormal modes (QNM) of FSCs arising from the singularity structure of our obtained formulae for the structure constants. Rather remarkably, we recover our very different answers for the singularity structure of these two types of BMS structure constants as the leading and subleading pieces of the BTZ QNM. We conclude in Sec.~6 with a summary of our results and some comments. 

\bigskip \bigskip

\section{Flat holography and BMS modular transformations}

The asymptotic symmetries of 3d asymptotically flat spacetimes at null infinity are given by the infinite-dimensional centrally extended BMS$_{3}$ algebra \eqref{cca}. The central message of the Carrollian version of holography, now adapted for 3d flat spacetimes, is that the dual field theory is a 2d theory that lives on the entire null boundary $(\mathscr{I}^{\pm})$, not just the circle at infinity and it inherits the underlying symmetry algebra given above \eqref{cca}. We call these theories 2d BMS invariant field theories or 2d BMSFTs for short. These 2d BMSFT defined on $\mathscr{I}^{\pm}$ also inherit the following metric
\begin{align}
ds^2=0 \times du^2+ d\theta^{2}.
\end{align}
Here $u$ is the retarded (null) time direction, and $\th$ is the angular direction of the celestial circle at null infinity. We now focus on a single BMSFT living on $\mathscr{I}^{+}$. The topology of the null boundary is that of a cylinder $\mathbb{R}_{u}\times S^{1}$. 

\medskip

Flat spacetime can be understood as an infinite radius\footnote{Or in other words a limit of vanishing comsmological constant.} limit of AdS. Hence it is natural to attempt a formulation of flat holography in terms of a limit of AdS/CFT where the AdS radius is taken to infinity. This singular limit in the bulk amounts to sending the speed of light to zero in the boundary CFT \cite{Bagchi:2012cy}, and results in what is called a Carrollian CFT. The isomorphism between the conformal Carroll algebra that is obtained as a contraction of the relativistic conformal algebra and the BMS algebra in one dimension higher \cite{Bagchi:2010zz, Duval:2014uva} is at the heart of this. The 2d Conformal Carroll or BMS$_3$ algebra can be obtained by an In\"on\"u-Wigner contraction of two copies of the Virasoro algebra 
\begin{align}
\label{IW_contractions}
    L_{n}=\mathcal{L}_{n}-\Bar{\mathcal{L}}_{-n},\quad M_{n}=\epsilon(\mathcal{L}_{n}+\Bar{\mathcal{L}}_{-n}),
\end{align}
where $\mathcal{L}_{n}$ and ${\bar{\mathcal{L}}}_n$ are the Virasoro generators. The parameter $\e$ can be thought of as the inverse of the AdS radius in the bulk and the speed of light on the boundary theory. 

\medskip

In the construction of Carroll holography for 3d asymptotically flat spacetimes, it is convenient to work with the highest-weight representation\footnote{Though also other representations have been considered previously in the literature. See, e.g., \cite{Campoleoni:2015qrh,Campoleoni:2016vsh,Oblak:2016eij}.} of the algebra \eqref{cca}. In this representation states 
 are labeled by the $L_{0}$ and $M_{0}$ eigenvalues, given by $\D$ and $\xi$ respectively \cite{Bagchi:2009ca}. We denote the states as $\ket{\Delta,\xi}$: 
\be{}
L_{0} \ket{\Delta,\xi}= \Delta \ket{\Delta,\xi}, \quad M_{0}\ket{\Delta,\xi}=\xi \ket{\Delta,\xi}. 
\ee
The action of $L_{n}$ and $M_{n}$ lowers the $\D$ eigenvalue of the states. If we take the spectrum of $\D$ to be bounded from below, there are states $\ket{\Delta_{p},\xi_{p}} \equiv \ket{\Delta,\xi}_{p} $ for which this eigenvalue cannot be lowered further and 
\begin{align}
L_{n}\ket{\Delta,\xi}_{p}=M_{n}\ket{\Delta,\xi}_{p}=0, \quad \forall n>0 . 
\end{align}
These states $\ket{\Delta,\xi}_{p}$ are called BMS primary states. A tower of descendant states can be obtained by acting on the state $\ket{\Delta,\xi}_{p}$ with $L_{-n}$ and $M_{-n}$ with $n>0$. The primary state $\ket{\Delta,\xi}_{p}$, the infinite tower of descendant states and the central charges $c_{L}$ and $c_{M}$ together make up a BMS module $\mathcal{B}(\Delta, \xi, c_{L}, c_{M})$. 

\subsection*{Modular properties of BMSFTs}
The partition function for a 2d BMSFT is given by 
\begin{align}
\label{partition_bms}
    Z_{\mathrm{BMS}} (\sigma,\rho)=\mathrm{Tr}\left[e^{2\pi i\sigma\left(L_{0}-\frac{c_{L}}{2}\right)}e^{2\pi i\rho\left(M_{0}-\frac{c_{M}}{2}\right)}\right]. 
\end{align}
For a 2d BMSFT defined on a null torus, a notion of modular transformations descends from the parent relativistic 2d CFT \cite{Bagchi:2012xr, Bagchi:2013qva}. This is given by 
\be{bmsmod}
    \sigma\to\frac{a\sigma+b}{c\sigma+d},\quad \rho\to\frac{\rho}{(c\s+d)^2}.
\ee
For the specific case of the $S$-transformation, which will be very useful for us below, this reads:
\be{}
S: \quad     (\sigma, \rho)\to\left(-\frac{1}{\sigma}, \frac{\rho}{\s^2}\right).
\ee
The above can be understood as an exchange of circles of the deformed torus, where one of the circles becomes infinitesimally small (but crucially does not completely collapse the torus to a circle).  

\medskip

We assume that the partition function $Z_{\mathrm{BMS}}$ (which is the torus zero-point function) is invariant under the BMS modular transformations \eqref{bmsmod} and exploit the invariance under the $S$-transformation to arrive at the BMS version of the Cardy formula:
\be{bmscardy}
    S_{\mathrm{BMS-Cardy}}= \log d(\D, \xi) = 2\pi \left(c_{L}\sqrt{\frac{2\xi}{c_{M}}}+\Delta\sqrt{\frac{c_{M}}{2\xi}}\right). 
\ee
In the above, $d(\D,\xi)$ is the asymptotic density of states. One can further exploit the modular covariance of the torus one-point function to arrive at a Cardy-like formula for the asymptotic structure constants of a 2d BMSFT 
\begin{align} 
\label{structure_constant_bms}
   C_{\Delta\mathcal{O}\Delta}\approx C_{\Delta_{\chi}\mathcal{O}\Delta_{\chi}} i^{-\Delta_{\mathcal{O}}}e^{\frac{\xi_{\mathcal{O}}}{2}\left(-\frac{\Delta}{\xi}+\frac{c_{L}}{c_{M}}\right)}
    e^{-2\pi\left(\sqrt{\frac{2\xi}{c_{M}}}\Delta_{\chi}+\sqrt{\frac{\xi}{2c_{M}}}\left(\frac{\Delta}{\xi}-\frac{c_{L}}{c_{M}}\right)\xi_{\chi}\right)} 
\end{align} 
In this formula above, $(\Delta,\xi)$, $(\Delta_{\chi},\xi_{\chi})$ and  $(\Delta_{\mathcal{O}},\xi_{\mathcal{O}})$ denote the weights of the heavy\footnote{Heavy (Light) in this context means that the BMS weights $\xi$ and $\Delta$ are large (small) in comparison to the central charges $c_M$ and $c_L$.} background, the lowest non-vacuum primary ${\chi}$ and a light probe ${\mathcal{O}}$ respectively {\footnote{We have taken the degeneracy of the vacuum and that of $\chi$ to be one here. Otherwise there would be a factor $\frac{d(\Delta_{\chi},\xi_{\chi})}{d(0,0)}$ on the right-hand-side.}}.

 \subsection*{Bulk implications}
 In the parlance of AdS$_3$/CFT$_2$, BTZ black holes are dual to thermal states in a 2d CFT. BTZ black holes are also orbifolds of AdS$_3$. There are similar orbifolds of 3d Minkowski spacetimes. Of particular interest are the shifted-boost orbifolds, which give rise to cosmological solutions called Flat Space Cosmologies (FSC) \cite{Cornalba:2002fi,Cornalba:2003kd}. These FSCs can be described by the following locally Ricci-flat metric \begin{equation}
\label{FSC_metric}
    ds^{2}=-\frac{dr^{2}}{f(r)}+f(r)^{2}dt^{2}+r^{2}(d\phi-N_{\phi}(r)dt)^{2},
\end{equation}
where 
\be{}
f(r)=\frac{\hat{r}_{+}^{2}(r^{2}-r_{0}^{2})}{r^{2}}, \quad N_{\phi}(r)=\frac{\hat{r}_{+}r_{0}}{r^{2}}.
\ee 
In the above, $r_{0}$ denotes the FSC horizon and $\hat{r}_+=\sqrt{8GM}$. The coordinates $-\infty<t<\infty$ and $0\leq\phi\leq2\pi$ are spacelike in the region outside of the horizon $r>r_0$. These objects can also be obtained in the flat limit ($\ell \to \infty$, where $\ell$ is the AdS radius) of a non-extremal BTZ. An intriguing aspect of the limit is that only the inner horizon of the BTZ $(r_{-})$ survives the limit, while the outer horizon $r_{+}$ gets pushed to infinity. More conretely,
\begin{align}
    r_{+}\to \ell\sqrt{8GM}=\ell \hat{r}_{+},\hspace{5mm} r_{-}\to r_{0}=\sqrt{\frac{2G}{M}}J,
\end{align}
where $M$ and $J$ are the mass and angular momentum of the FSC. Like a BTZ black holes, an FSC can be associated with having a Hawking temperature $T_{FSC}=\frac{\hat{r}_{+}^{2}}{2\pi r_{0}}$. The BMS-Cardy entropy can be shown to exactly match the entropy of the FSC horizon \cite{Bagchi:2012xr,Bagchi:2013qva,Bagchi:2019unf}. In the dual bulk theory, the three-point coefficient in BMSFT is interpreted as the expectation value of a probe operator in the background FSC states \cite{Bagchi:2020rwb}. 
 

\section{Off-Diagonal Structure Constant in BMSFT}
By exploiting modular covariance of BMSFTs on a torus, it is possible to derive the asymptotic density of states and an asymptotic formula for structure constants from the zero-point (partition function) and one-point functions on the torus. It is natural to take this line of inquiry forward and explore the implications of what physics the torus two-point function can elucidate. In the context of AdS$_3$/CFT$_2$, the analogous analysis provides some interesting insights into various phenomena, including the eigenstate thermalization hypothesis and the BTZ black hole quasinormal mode spectrum. With the hope of connecting to similar phenomena in the context of flat holography, we now delve into the details of this computation. 

\bigskip

Using the modular covariance of BMS two-point functions we derive the off-diagonal structure constants in 2D BMSFT. We start with the expression of a BMS two-point function on the cylinder:
\begin{equation}
    \langle\mathcal{O}(u_{1},\theta_{1})\mathcal{O}(u_{2},\theta_{2})\rangle = {\sin\left(\frac{\theta_{1}-\theta_{2}}{2}\right)^{-2\Delta_{\mathcal{O}}}}\exp\left[-\xi_{\mathcal{O}}(u_{1}-u_{2})\cot\left(\frac{\theta_{1}-\theta_{2}}{2}\right)\right].
\end{equation}
Here $u$ and $\theta$ are the (null) time and space coordinates on the cylinder that describes e.g. $\mathscr{I}^+$ in asymptotically flat 3d spacetime. To get to the torus, we identify the ends of the cylinder. We have already introduced the modified modular transformations for 2d BMSFTs. This is given by \eqref{bmsmod}. The elliptic coordinates $(u, \th)$ of the BMS torus also change under modular transformations: 
\begin{equation}
    u\to\frac{u}{c\sigma+d}-\frac{\theta\rho c}{(c\sigma+d)^{2}},\quad \theta\to\frac{\theta}{c\sigma+d}.
\end{equation}
In a 2d BMSFT, a BMS transformation from a set of coordinates $(x, y) \to (x^{\prime}, y^{\prime})$ leads to a transformation on BMS primaries: 
 \begin{align}
     \Tilde{\mathcal{O}}(x^{\prime},y^{\prime})=(\partial_{y}y^{\prime})^{-\Delta} \exp{\xi\frac{\partial_{y}x^{\prime}}{\partial_{y}y^{\prime}}}\mathcal{O}(x,y).   
 \end{align}
This means that under BMS transformations, the two-point function of primaries transforms as:
\begin{eqnarray}
\label{trafo1}
\left\langle\mathcal{O}_{1}(x^{\prime}_{1},y^{\prime}_{1})\mathcal{O}_{2}(x^{\prime}_{2},y^{\prime}_{2})\right\rangle&=&
\prod_{i=1}^2 \left[(\partial_{y}y^{\prime})^{-\Delta_{i}}\exp(\xi_{i}\frac{\partial_{y}x^{\prime}}{\partial_{y}y^{\prime}})\right]_{\substack{x=x_{i}\\y=y_{i}}} \left\langle\mathcal{O}_{1}(x_{1},y_{1})\mathcal{O}_{2}(x_{2},y_{2})\right\rangle
\end{eqnarray}
We express the two-point function on the torus with modular parameters $\sigma$ and $\rho$ as a trace taken over highest-weight states and their descendants: 
\begin{eqnarray}\label{3.5}
  \langle\mathcal{O}(u_{1},\theta_{1})\mathcal{O}(u_{2},\theta_{2})\rangle_{(\sigma,\rho)}&=&\mathrm{Tr}\left[\mathcal{O}(u_{1},\theta_{1})\mathcal{O}(u_{2},\theta_{2})e^{2\pi i\sigma(L_{0}-c_{L}/2)}e^{2\pi i\rho(M_{0}-c_{M}/2)}\right]\nonumber\\ 
  &=&\sum_{i,j} e^{2\pi i\sigma(\Delta_{i}-c_{L}/2)}e^{2\pi i\rho(\xi_{i}-c_{M}/2)}|\langle\Delta_{i},\xi_{i}|\mathcal{O}(0,0)|\Delta_{j},\xi_{j}\rangle|^{2} e^{i (u_{12} \xi_{ij} +\theta_{12}\Delta_{ij})}\nonumber\\
 &=& \widetilde{\langle\mathcal{O}(u_{1},\theta_{1})\mathcal{O}(u_{2},\theta_{2})\rangle}_{(\sigma,\rho)} e^{-2\pi i(\sigma c_{L}/2+\rho c_{M}/2)}.
 \end{eqnarray}
In the second step, the trace is taken over all primaries and their descendants. We have also inserted a complete set of states, $\sum_{j}\ket{\Delta_{j},\xi_{j}}\bra{\Delta_{j},\xi_{j}}$. Finally, time evolution and spatial translation of the operator $\O$ is given by
\be{}
\mathcal{O}(u,\theta)= e^{i H u+i P\theta}\mathcal{O}(0,0)e^{-i H u-i P\theta}= e^{i M_0 u+i L_0\theta}\mathcal{O}(0,0)e^{-i M_0 u-i L_0\theta},
\ee 
where we have used $H=i\partial_{u}=M_{0}$ and $P=i\partial_{\theta}=L_{0}$. Notice the subscript on the two-point function on the l.h.s of the equation \eqref{3.5}. Here $(\s,\rho)$ denotes the modular parameters which we will keep track of. In the last step of \eqref{3.5}, the $\widetilde{\<\O \O\>}$ is just $\<\O \O\>$ with the central terms stripped off: 
\begin{equation}
\widetilde{ \langle\mathcal{O}(u_{1},\theta_{1})\mathcal{O}(u_{2},\theta_{2})\rangle}_{(\sigma,\rho)}=\sum_{i,j} e^{2\pi i\sigma\Delta_{i}}e^{2\pi i\rho\xi_{i}}|\langle\Delta_{i},\xi_{i}|\mathcal{O}(0,0)|\Delta_{j},\xi_{j}\rangle|^{2} e^{i (u_{12} \xi_{ij} +\theta_{12}\Delta_{ij})}.   
\end{equation}
In the above, we have used abbreviations like $u_{12}=u_1-u_2$. Changing the summation to integration and then inverting the integral we can extract the off-diagonal three-point coefficients, which can be obtained upon solving
  \begin{equation}      d(\Delta,\xi)d(\Delta^{\prime},\xi^{\prime})\overline{|C_{\Delta\xi\mathcal{O}\Delta^{\prime}\xi^{\prime}}|^{2}}=\int \mathrm{d}\sigma\mathrm{d}\rho\mathrm{d}u\mathrm{d}\theta \, e^{-2\pi i(\sigma c_{L}+\rho c_{M})}e^{-i(u_{12}\zeta+\theta_{12}\lambda)} \widetilde{ \langle\mathcal{O}(u_{1},\theta_{1})\mathcal{O}(u_{2},\theta_{2})\rangle}_{(\sigma,\rho)},
  \end{equation}
 where we have introduced 
 \be{}
 \zeta=\xi-\xi^{\prime} \quad \lambda=\Delta-\Delta^{\prime},
 \ee 
and $d(\D,\xi)$ and $d(\D',\xi')$ are the density of states with weights $(\D, \xi)$ and $(\D', \xi')$ respectively and $d(\D',\xi')$ is given by
\be{}
d(\D,\xi) = \exp{2\pi \left(c_{L}\sqrt{\frac{2\xi}{c_{M}}}+\Delta\sqrt{\frac{c_{M}}{2\xi}}\right)}.
\ee
\medskip 
Now, using the BMS transformation behavior of the two-point function one can relate $\widetilde{\langle\mathcal{O}(u_{1},\theta_{1})\mathcal{O}(u_{2},\theta_{2})\rangle}_{(\sigma,\rho)}$ with its $S$-modular transformed counterpart. The S-modular transformation on the elliptic coordinates reads 
\be{moduth}
  S: \quad (u, \theta) \to (u^{\prime}, \theta^{\prime}) = \left(\frac{u}{\sigma}-\frac{\theta\rho}{\sigma^{2}}, \frac{\theta}{\sigma} \right).   
\end{equation}
 We now use the change in the two-point function under BMS modular $S$-transformation: 
\begin{equation}
\label{S-Modular_2pt}
    \widetilde{\langle\mathcal{O}(u_{1},\theta_{1})\mathcal{O}(u_{2},\theta_{2})\rangle}_{(\sigma,\rho)}=\sigma^{-2\Delta_{\mathcal{O}}}e^{\frac{2\xi_{\mathcal{O}}\rho}{\sigma}} e^{\pi i\left(\sigma c_{L}+\rho c_{M}+\frac{c_{L}}{\sigma}-\frac{\rho c_{M}}{\sigma^{2}}\right)}\widetilde{\langle\mathcal{O}(u_{1}^{\prime},\theta_{1}^{\prime})\mathcal{O}(u_{2}^{\prime},\theta_{2}^{\prime})\rangle}_{\left(-\frac{1}{\sigma},\frac{\rho}{\sigma^{2}}\right)}.
\end{equation}
The modular parameters $\sigma$ and $\tau$ are related to the twist of the torus $\Omega$ and temperature $\beta$ through the following relations:
\begin{equation}
    2\pi\sigma=i\Omega\quad,\quad 2\pi\rho=\beta.
\end{equation}
This $S$-transformation takes $\Omega\to -\frac{1}{\Omega}$. Thus the vacuum state dominates the $S$-transformed correlation function since the contributions from non-vacuum states are exponentially suppressed at low values of the twist. Inserting the expression of the cylinder two-point function into \eqref{S-Modular_2pt} one obtains
\begin{eqnarray}
     \widetilde{\langle\mathcal{O}(u_{1},\theta_{1})\mathcal{O}(u_{2},\theta_{2})\rangle}_{(\sigma,\rho)}&=&\sigma^{-2\Delta_{\mathcal{O}}}e^{2\xi_{\mathcal{O}}\rho/\sigma} e^{2\pi i\left(\frac{\sigma c_{L}}{2}+\frac{\rho c_{M}}{2}+\frac{c_{L}}{2\sigma}-\frac{\rho c_{M}}{2\sigma^{2}}\right)} \left[\sin\left(\frac{\theta'_{12}}{2}\right)\right]^{-2\Delta_{\mathcal{O}}}\nonumber\\
     &&\quad \times \exp\left[-\xi_{\mathcal{O}}u'_{12}\cot\left(\frac{\theta'_{12}}{2}\right)\right],
\end{eqnarray}
where $u', \theta'$ are given by the modular transformed $u, \theta$ in \eqref{moduth}. One now has the following integration to solve to get the expression for $\overline{|C_{\Delta\xi\mathcal{O}\Delta^{\prime}\xi^{\prime}}|^{2}}$:
\begin{eqnarray}
\label{starting_eqn}
d(\Delta,\xi)d(\Delta^{\prime},\xi^{\prime})\overline{|C_{\Delta\xi\mathcal{O}\Delta^{\prime}\xi^{\prime}}|^{2}}&=&\int \mathrm{d}\sigma\mathrm{d}\rho\mathrm{d}u\mathrm{d}\theta \, e^{-2\pi i(\sigma c_{L}+\rho c_{M})}e^{-i(u'_{12}\zeta+\theta'_{12}\lambda)} \sigma^{-2\Delta_{\mathcal{O}}}e^{2\xi_{\mathcal{O}}\rho/\sigma}\nonumber\\
  &&e^{2\pi i\left(\frac{\sigma c_{L}}{2}+\frac{\rho c_{M}}{2}+\frac{c_{L}}{2\sigma}-\frac{\rho c_{M}}{2\sigma^{2}}\right)}\frac{\exp\left[-\xi_{\mathcal{O}}u'_{12}\cot\left(\frac{\theta'_{12}}{2}\right)\right]}{ \left[\sin\left(\frac{\theta'_{12}}{2}\right)\right]^{2\Delta_{\mathcal{O}}}}.
\end{eqnarray}
The above integral is very difficult to solve in full generality and its solution gives the most general off-diagonal three-point function coefficients in a BMSFT. To provide tractable solutions we focus on two specific cases, one where the probes are only temporally separated and one where the probes are only spatially separated. The temporally separated case is analog to the 2d CFT case and is an indicator for ETH in 2d CFTs. We will elaborate on this in detail in the next section. The spatially separated case for BMSFTs has very interesting features which we will also comment on in the next section. But before we move on to the two cases, we do a few simplifications to the above equation \eqref{starting_eqn}. First, we focus on Einstein gravity as the dual bulk theory, so we put $c_L=0$. We also position the probes at $(u',\theta')$ and $(0,0)$ without loss of generality. Inserting the modular transformed $(u',\theta')$ in terms of $(u, \theta)$, \eqref{starting_eqn} now simplifies to:
\begin{eqnarray}
\label{starting_eqn1}
  d(\Delta,\xi)d(\Delta^{\prime},\xi^{\prime})\overline{|C_{\Delta\xi\mathcal{O}\Delta^{\prime}\xi^{\prime}}|^{2}}&=&\int \mathrm{d}\sigma\mathrm{d}\rho\mathrm{d}u\mathrm{d}\theta \, e^{-2\pi i \rho c_{M}}e^{-i((\frac{u}{\s}+\frac{\th \rho}{\s^2})\zeta+\frac{\theta}{\s}\lambda)} \sigma^{-2\Delta_{\mathcal{O}}}e^{2\xi_{\mathcal{O}}\rho/\sigma} e^{\pi i \rho c_{M}\left(1 -\frac{1}{\sigma^{2}}\right)} \nonumber\\
  &&\quad \sin\left(\frac{\theta}{2\s}\right)^{-2\Delta_{\mathcal{O}}} {\exp\left[-\xi_{\mathcal{O}} \left(\frac{u}{\s}-\frac{\th \rho}{\s^2}\right) \cot\left(\frac{\theta}{2\s}\right)\right]}.
\end{eqnarray}

\bigskip

\bigskip

\subsection{Temporally Separated Probes}
We first consider the case where the probes are temporally separated. To make sure that there are no divergences in our computations, we start with $\langle\D^{\prime}, \xi^{\prime}|\mathcal{O}(u,\th_c)\mathcal{O}(0,0)|\D, \xi\rangle$ and finally take $\th_c\to 0$. Also, for this calculation, we take $\D=\D'$, i.e. the initial and final angular momenta of the heavy background are the same.

\medskip

At equal $\D$, there is no $\th$ integration and \eqref{starting_eqn1} reduces to: 
\begin{align}
    \label{saddle_bms}
    d(\Delta,\xi) d(\Delta,\xi^{\prime})\overline{|C_{\xi\mathcal{O}\xi^{\prime}}|^{2}}=   \int & \mathrm{d}u \ e^{-i u\zeta} \int\mathrm{d}\sigma\mathrm{d}\rho\hspace{2mm}\sigma^{-2\Delta_{\mathcal{O}}}e^{2\xi_{\mathcal{O}}\rho/\sigma}e^{2\pi i\left(\frac{\rho c_{M}}{2}-\frac{\rho c_{M}}{2\sigma^{2}}\right)} e^{-2\pi i\sigma\Delta}e^{-2\pi i\rho\xi}\nonumber\\ 
   & {\left(\sin{\frac{\theta_{c}}{2\sigma}}\right)^{-2\Delta_{\mathcal{O}}}} \exp{-\xi_{\mathcal{O}}\left(\frac{u}{\sigma}-\frac{\rho \theta_{c}}{\sigma^{2}}\right)\cot\left(\frac{\theta_{c}}{2\sigma}\right)}.
     \end{align}
We first perform the integration over $\s$ and $\rho$. Since the background is much heavier than the probe $\mathcal{O}$, one can perform a saddle-point approximation. Computing the saddle points yields $(\sigma,\rho)$: 
\begin{align}
    \sigma_{c}=i\sqrt{\frac{c_{M}}{2\xi}}, \quad \rho_{c}=-i\frac{\Delta\sqrt{c_{M}}}{(2\xi)^{3/2}}.
\end{align}
As $\xi\to\infty$, one can approximate 
\begin{equation}
    \sinh\left(\frac{\theta_{c}\sqrt{\xi}}{\sqrt{2 c_{M}}}\right)\to e^{\theta_{c}\sqrt{\xi}/\sqrt{2 c_{M}}}, \quad \coth\left(\frac{\theta_{c}\sqrt{\xi}}{\sqrt{2 c_{M}}}\right)\to 1.
\end{equation}
Since $\theta_{c}\to 0$, we require that $\sqrt{\frac{\xi}{2c_{M}}}$ goes to infinity faster than $\theta_{c}\to0$. Putting this in equation \ref{saddle_bms} one obtains,
\begin{align}
 d(\Delta,\xi)d(\Delta,\xi^{\prime})\overline{|C_{\xi\mathcal{O}\xi^{\prime}}|^{2}}&\approx f(\Delta,\xi) \int_{0}^{\infty} e^{-\xi_{\mathcal{O}}u\sqrt{\frac{2\xi}{c_{M}}}} e^{-i u\zeta} \mathrm{d}u =\frac{f(\Delta,\xi)}{\left(i\zeta+\xi_{\mathcal{O}}\sqrt{\frac{2\xi}{c_{M}}}\right)},
 \end{align}
where the prefactor 
\be{}
f(\Delta,\xi)=\left(\frac{\xi}{c_{M}}\right)^{\Delta_{\mathcal{O}}}e^{-\xi_{\mathcal{O}}\Delta/\xi}e^{2\pi\left(\frac{\Delta}{2}\left(\frac{c_{M}}{2\xi}\right)^{3/2}+\Delta\sqrt{\frac{c_{M}}{2\xi}}\right)} e^{- \theta_{c} {\sqrt{\frac{2\xi}{c_M}}} \left(\Delta_\O + \frac{\xi_\O \Delta}{2\xi} \right)}.
\ee 
We stress that this is a probe approximation, i.e., we work in the limit when $\D_\O, \xi_\O \ll \xi$ {\footnote{Even though we previously assumed $\sqrt{\frac{\xi}{2c_{M}}}$ goes to infinity faster than $\theta_{c}\to0$, the additional factor containing the weight of the probes justifies our result.}. This means one can drop the last term in the above expression.   
Inserting the expressions for the density of states one gets 
\begin{equation}
\label{same_delta}
  \overline{|C_{\xi\mathcal{O\xi^{\prime}}}|^{2}}\approx \left(\frac{\xi_{avg}}{c_{M}}\right)^{\Delta_{\mathcal{O}}} e^{-\xi_{\mathcal{O}}\frac{\Delta_{avg}}{\xi_{avg}}}e^{-2\pi\Delta_{avg}\sqrt{\frac{c_{M}}{2\xi_{avg}}}\left(1-\frac{c_{M}}{4\xi_{avg}}\right)} \frac{1}{\left(i\zeta+\xi_{\mathcal{O}}\sqrt{\frac{2\xi_{avg}}{c_{M}}}\right)}.  
\end{equation}
In the above, we have replaced the weights of the states with average weights given by 
\begin{align}
    \Delta+\Delta^{\prime}=2\Delta=2\Delta_{\mathrm{avg}},\quad\quad \xi+\xi^{\prime}=2\xi_{\mathrm{avg}},
\end{align}
where we have neglected terms $\mathcal{O}\left(\frac{\zeta}{\xi_{\mathrm{avg}}}\right)$ since $\xi_{\mathrm{avg}}\gg\zeta$.  

\bigskip

\subsection{Spatially Separated Probes}
 We now investigate the case of spatially separated probes. The operators are now inserted at $(0,\th)$ and $(0,0)$. Thus, one is looking at correlations of the form $\langle\D^{\prime}, \xi^{\prime}|\mathcal{O}(0,\th)\mathcal{O}(0,0)|\D, \xi\rangle$. We shall also take $\xi=\xi'$. After carrying out the saddle point analysis, one is now left with 
\begin{align}
 d(\Delta,\xi)d(\Delta^{\prime},\xi) \overline{|C_{\Delta\mathcal{O}\Delta^{\prime}}|^{2}} \approx g(\Delta,\xi)
\int \mathrm{d}\theta \, \frac{e^{-i\theta\lambda} e^{-\xi_{\mathcal{O}}\frac{\theta\Delta}{\sqrt{2c_{M}\xi}}\coth{\theta\sqrt{\frac{\xi}{2c_{M}}}}}}{\left(\sinh\theta\sqrt{\frac{\xi}{2c_{M}}}\right)^{2\Delta_{\mathcal{O}}}},
\end{align}
where 
\begin{align}
 g(\Delta,\xi)=e^{-\xi_{\mathcal{O}}\Delta/\xi} \left(\frac{\xi}{c_{M}}\right)^{\Delta_{\mathcal{O}}} \exp{2\pi\left(\Delta\sqrt{\frac{c_{M}}{2\xi}}+\frac{\Delta}{2}\left(\frac{c_{M}}{2\xi}\right)^{3/2}\right)}.   
\end{align}
Since the background field is very heavy and hence $\xi\gg1$, so we approximate $\coth\left({\theta\sqrt{\frac{\xi}{2c_{M}}}}\right)$ by 1. However, in this case we do not approximate $\sinh\left(\frac{\theta\sqrt{\xi}}{\sqrt{2 c_{M}}}\right)\to e^{\theta\sqrt{\xi}/\sqrt{2 c_{M}}}$. {\footnote{This is done in order to be able to make use of certain Mellin-Barnes integral identities, as we show in what follows.}} We also work with de-compactified spatial coordinates. One now has,
\begin{align}
d(\Delta,\xi)d(\Delta^{\prime},\xi)\overline{|C_{\Delta\mathcal{O}\Delta^{\prime}}|^{2}}&\approx g(\Delta,\xi) \int_{-\infty}^{\infty} \mathrm{d}\theta \left(\sinh\theta\sqrt{\frac{\xi}{2c_{M}}}\right)^{-2\Delta_{\mathcal{O}}} \exp{-i\theta\left(\frac{\xi_{\mathcal{O}}\Delta}{i\sqrt{2c_{M}\xi}}+\lambda\right)}  \nonumber\\
=g(\Delta,\xi) & \sqrt{\frac{c_{M}}{\xi}}\exp{\pi\lambda\sqrt{\frac{c_{M}}{2\xi}}}\frac{\bigg|\Gamma\left(\Delta_{\mathcal{O}}+\frac{\xi_{\mathcal{O}}\Delta}{2\xi}+i\sqrt{\frac{c_{M}}{2\xi}}\lambda\right)\bigg|^{2}}{\Gamma(2\Delta_{\mathcal{O}})}.
\end{align} 
The $\theta$ integration can be solved by using the Mellin-Barns integral identity (see, e.g., \cite{Jantzen:2012cb,Becker:2014jla}). Writing $\overline{|C_{\Delta\mathcal{O}\Delta^{\prime}}|^{2}}$ in terms of $\Delta_{avg}$ and neglecting terms of $\mathcal{O}(\lambda/\Delta_{avg})$ we find,
\begin{eqnarray}
\overline{|C_{\Delta\mathcal{O}\Delta^{\prime}}|^{2}}
 &\approx& e^{-\xi_{\mathcal{O}}\frac{\Delta_{avg}}{\xi_{avg}}}\left(\frac{\xi_{avg}}{c_{M}}\right)^{\Delta_{\mathcal{O}}-1/2} e^{-2\pi\Delta_{avg}\sqrt{\frac{c_{M}}{2\xi_{avg}}}\left(1-\frac{c_{M}}{4\xi_{avg}}-\frac{\lambda}{2\Delta_{avg}} \right)} e^{-i\pi\frac{\xi_{\mathcal{O}}\Delta_{avg}}{2\xi_{avg}}}\nonumber\\
 &&\qquad\qquad\qquad\qquad\frac{\bigg|\Gamma\left(\Delta_{\mathcal{O}}+\frac{\xi_{\mathcal{O}}\Delta_{avg}}{2\xi_{avg}}+i\sqrt{\frac{c_{M}}{2\xi_{avg}}}\lambda\right)\bigg|^{2}}{\Gamma(2\Delta_{\mathcal{O}})}.
\end{eqnarray}
With $\frac{\lambda}{\Delta_{avg}}\ll1$, one can omit terms $\mathcal{O}\left(\frac{\lambda}{\Delta_{\mathrm{avg}}}\right)$ in the exponential. Dropping $e^{-i\pi\frac{\xi_{\mathcal{O}}\Delta_{avg}}{2\xi_{avg}}}$ in the probe approximation, we finally arrive at the expression for the off-diagonal three-point coefficient for spatially separated probes: 
\begin{equation}
\label{same_xi}
 \overline{|C_{\Delta\mathcal{O}\Delta^{\prime}}|^{2}}
 \approx \left(\frac{\xi_{avg}}{c_{M}}\right)^{\Delta_{\mathcal{O}}-1/2} e^{-\xi_{\mathcal{O}}\frac{\Delta_{avg}}{\xi_{avg}}} e^{-2\pi\Delta_{avg}\sqrt{\frac{c_{M}}{2\xi_{avg}}}\left(1-\frac{c_{M}}{4\xi_{avg}}\right)}
 \frac{\bigg|\Gamma\left(\Delta_{\mathcal{O}}+\frac{\xi_{\mathcal{O}}\Delta_{avg}}{2\xi_{avg}}+i\sqrt{\frac{c_{M}}{2\xi_{avg}}}\lambda\right)\bigg|^{2}}{\Gamma(2\Delta_{\mathcal{O}})}.   
\end{equation}

\bigskip

\section{Eigenstate Thermalization Hypothesis}
The eigenstate thermalization hypothesis provides us with a mechanism to study the late-time behavior of a thermal system, i.e., thermalization. In the observable dependent notion of thermalization, one studies the coarse-grained expectation value of a generic observable $\mathcal{O}$ over the states $\ket{\psi}$. The states $\ket{\psi}$ are a superposition of the energy eigenstates of the quantum system and are given by $\ket{\psi}=\sum_{n}C_{n}\ket{n}$. The expectation value of $\mathcal{O}$ at time $t$ is given by, 
\begin{align}
    \langle\mathcal{O}(t)\rangle=\sum_{n}\vert C_{n}\vert^{2}\bra{n}\mathcal{O}\ket{n}+\sum_{n,m\neq n} C_{n}^{*}C_{m}e^{i(E_{n}-E_{m})t}\bra{n}\mathcal{O}\ket{m},
\end{align}
where $E_{n}$ denotes the energy of the state $\ket{n}$. For a system with a large number of degrees of freedom, the late time average of $\mathcal{O}(t)$ is equivalent to the thermal average of the operator \cite{Srednicki:1995pt}. The late time average is given by, 
\begin{align}
    \overline{\langle\mathcal{O}(t)\rangle}=\lim_{\mathcal{T}\to\infty}\frac{1}{\mathcal{T}}\int_{0}^{\mathcal{T}} \mathrm{d}t \langle\mathcal{O}(t)\rangle= \vert C_{n}\vert^{2} \bra{n}\mathcal{O}\ket{n}.  
\end{align} 
The fluctuations of the system are given by the difference between the expectation value of the observable and its late time average. They are encoded in the off-diagonal part of the expectation value. We have
\begin{align}
    \langle\mathcal{O}(t)\rangle-\overline{\langle\mathcal{O}(t)\rangle}= \sum_{m\neq n} C_{n}^{*}C_{m}e^{i(E_{n}-E_{m})t}\bra{n}\mathcal{O}\ket{m}.  
\end{align}
For the system to thermalize, this fluctuation has to decay very fast or become reasonably negligible. ETH states that the off-diagonal element $\bra{n}\mathcal{O}\ket{m}$ fall of as $e^{-S(E_{\mathrm{avg}})/2}$ where $S(E_{\mathrm{avg}})$ is the entropy of the system at the average energy $E_{\mathrm{avg}}=\frac{E_{n}+E_{m}}{2}$. 

\subsection{Structure Constants for 2d CFT}
Our BMS analysis has been inspired by the calculations in 2d CFTs \cite{Brehm:2018ipf, Hikida:2018khg,Romero-Bermudez:2018dim}. Here we first recount briefly the 2d CFT result and then the physics that can be extracted from this answer. A 2d CFT has well-known modular properties. The partition function 
\be{}
Z_{\text{CFT}}= {\text{Tr}} \, e^{2\pi i \t (\L_0+\frac{c}{24})} e^{-2\pi i \bar{\t} (\bar{\L}_0 +\frac{\bar{c}}{24})},
\ee
is invariant under modular transformations that take the modular parameter $(\t, \bar{\t})$ and the elliptic coordinates on the torus $(\w, \bar{\w})$ to 
\begin{equation}
  \tau\to\frac{a\tau+b}{c\tau+d}\quad,\quad w\to\frac{w}{c\tau+d},   
\end{equation}
and similarly for the barred components. The correlators on the torus transform covariantly under the modular group. In particular, the two-point function transforms as
\begin{equation}
    \langle\mathcal{O}(w_{1},\Bar{w}_{1})\mathcal{O}(w_{2},\Bar{w}_{2})\rangle_{\tau}=(c\tau+d)^{-2h_{\mathcal{O}}}(c\Bar{\tau}+d)^{-2\Bar{h}_{\mathcal{O}}} \langle\mathcal{O}(w_{1}^{\prime},\Bar{w}_{1}^{\prime})\mathcal{O}(w_{2}^{\prime},\Bar{w}_{2}^{\prime})\rangle_{\frac{a\tau+b}{c\tau+d}} .
\end{equation}
We are interested in the autocorrelations $\langle\mathcal{O}(t,0)\mathcal{O}(0,0)\rangle_{\beta}$, where $\beta$ is the inverse temperature which is related to the modular parameter by $\beta = i(\t - \bar{\t})$. One can exploit the $S$-modular covariance of the two-point function, do a saddle-point analysis, and arrive at a formula for the off-diagonal structure constants, similar to what we have done in the previous section for BMSFTs. The off-diagonal structure constants for a 2d CFT are given by 
\begin{eqnarray} 
\label{gamma_pole}
    \overline{|C_{E\mathcal{O}E^{\prime}}|^{2}}\sim e^{-S_{avg}}\left(\frac{12E_{avg}}{c}-1\right)^{E_{\mathcal{O}}+1/4}\bigg|\Gamma\left(E_{\mathcal{O}}+i\frac{\omega}{\sqrt{12E_{avg}/c-1}}\right)\bigg|^{2}.
\end{eqnarray}
In the above, we have set the central terms to be equal $c=\bar{c}$ as is the case for Einstein gravity. We have defined $E=h+\h$ and $\omega=E^{\prime}-E$. Also $E_{avg}=\frac{1}{2}(E^{\prime}+E)$. $S_{avg}$ is given by the Cardy formula $$S_{avg} = 2\pi \sqrt{\frac{c E_{avg}}{3}}.$$
Finally, in \eqref{gamma_pole}, $E_\O$ is the weight of the probe $\O$. It is clear from the above calculation that the off-diagonal matrix elements in a 2d CFT are suppressed as $e^{-S_{avg}}$, and hence 2d CFTs follow ETH and thermalize. 

\subsection{ETH and BMS}
As highlighted earlier, our main results in this work are expressions for different types of BMS structure constants given by Eq.~\eqref{same_delta} and Eq.~\eqref{same_xi}. 

\medskip

\paragraph{Temporally separated probes.}
We first focus on the answer for temporally separated probes: 
\begin{equation}
\label{eqD}
\overline{|C_{\xi\mathcal{O\xi^{\prime}}}|^{2}}\approx \left(\frac{\xi_{avg}}{c_{M}}\right)^{\Delta_{\mathcal{O}}} e^{-\xi_{\mathcal{O}}\frac{\Delta_{avg}}{\xi_{avg}}}e^{-2\pi\Delta_{avg}\sqrt{\frac{c_{M}}{2\xi_{avg}}}\left(1-\frac{c_{M}}{4\xi_{avg}}\right)} \frac{1}{\left(i\zeta+\xi_{\mathcal{O}}\sqrt{\frac{2\xi_{avg}}{c_{M}}}\right)}.  
\end{equation}
This is the analog of the 2d CFT analysis that we presented previously. Notice that 
\be{}
\exp{-2\pi\Delta_{\mathrm{avg}}\sqrt{\frac{c_{M}}{2\xi_{avg}}}} =  \exp{-S^{\text{BMS}}_{avg}},
\ee
when the central charge $c_L=0$
. This corresponds to asymptotically flat spacetimes in 3d Einstein gravity as a dual bulk theory. However, the exponentially suppressed prefactor $p(\D_{avg}, \xi_{avg}, c_M)$ comes with another piece: 
\be{p}
p(\D_{avg}, \xi_{avg}, c_M) = \exp{-S^{\text{BMS}}_{avg}\left(1-\frac{c_{M}}{4\xi_{avg}}\right)}.
\ee
We are in the Cardy regime, meaning that $\xi_{avg}\gg c_M$. So, $\frac{c_{M}}{4\xi_{avg}} \ll 1$ in this region of parameter space. Hence the off-diagonal structure constants are indeed exponentially suppressed compared to the diagonal entries. This suppression, however, is not completely analogous to a 2d CFT because of this additional piece. We take this to be a signature of the non-Lorentzian nature of the theory. The fact that there is exponential suppression means that the system seems to follow ETH and thermalizes. One can also infer from the change in the exponential structure that the thermalization is now slower compared to the 2d CFT case. 

\medskip

Before we move on to discuss the other case, it is important to mention that there are additional prefactors on the r.h.s of \eqref{eqD} multiplying the exponential suppression. We will deal with the singularity structure arising from the last term extensively in the next section. For now, we comment on the first couple of terms:  
\be{s1}
s_1 =\left(\frac{\xi_{avg}}{c_{M}}\right)^{\Delta_{\mathcal{O}}} e^{-\xi_{\mathcal{O}}\frac{\Delta_{avg}}{\xi_{avg}}}.
\ee
These terms are interpreted as probe dependent sub-leading corrections to the entropy of the FSC in the bulk, as they explicitly depend on the weights of the probe $\O$. 

\bigskip

\medskip

\paragraph{Spatially separated probes.}
Let us now consider spatially separated probes. The answer for the off-diagonal structure constants in this case is given by 
\bea{eqxi}
\overline{|C_{\Delta\mathcal{O}\Delta^{\prime}}|^{2}}
 \approx  \left(\frac{\xi_{avg}}{c_{M}}\right)^{\Delta_{\mathcal{O}}-1/2} && e^{-\xi_{\mathcal{O}}\frac{\Delta_{avg}}{\xi_{avg}}} e^{-2\pi\Delta_{avg}\sqrt{\frac{c_{M}}{2\xi_{avg}}}\left(1-\frac{c_{M}}{4\xi_{avg}}\right)} \cr 
&&\frac{\bigg|\Gamma\left(\Delta_{\mathcal{O}}+\frac{\xi_{\mathcal{O}}\Delta_{avg}}{2\xi_{avg}}+i\sqrt{\frac{c_{M}}{2\xi_{avg}}}\lambda\right)\bigg|^{2}}{\Gamma(2\Delta_{\mathcal{O}})}.   
\eea
One can see that the same exponential suppression of the off-diagonal structure constants with the prefactor $p(\D_{avg}, \xi_{avg}, c_M)$ occurs. Even though the answer here does not a priori have a relation to late-time thermalization of eigenstates, the structure of the answer is interesting. An important point to note is that the bulk dual is an FSC solution that comes from a limit of a non-extremal BTZ black hole. Here the outer radius is pushed out to infinity and one is left with the inside of the black hole where the roles of time and space reverse. The dual field theory should also carry signatures of this flipping of the temporal and spatial direction. The indication of the answer above is similar. This is also reminiscent of the computations of chaos in Carrollian CFTs in \cite{Bagchi:2021qfe}, where one found trajectories diverge with Lyapunov coefficients while looking at spatial evolution instead of temporal ones. We thus interpret the exponential suppression in the above as a spatial analog of ETH and also thermalization. 

\medskip

In $d=2$ there is an interesting duality between the $c\to0$ Carrollian theories and the $c\to\infty$ Galilean theories. For non-Lorentzian CFTs, there is an isomorphism between the Carrollian and Galilean conformal algebras. This isomorphism effectively exchanges the identification of the temporal and spatial directions. The analysis we have performed in the 2d BMSFT is mostly agnostic to whether the theory came from a Galilean or a Carrollian limit. We have used BMS modular invariance, which also holds in a 2d Galilean CFT. The only thing that changes is the identification of space and time. So, the analysis done in the above sections is equally applicable for 2d Galilean CFTs, albeit with a flip of temporal and spatial coordinates. The answer \eqref{eqxi} is thus what one would get for off-diagonal structure constants in the case of temporally separated probes of 2d Galilean CFTs. The presence of the exponentially suppressing prefactor $p(\D_{avg}, \xi_{avg}, c_M)$ indicates that even for 2d Galilean CFTs, there is a notion of thermalization for late times in terms of ETH.  

\medskip

Before closing this section, we point out the additional prefactors on the r.h.s of \eqref{eqxi} multiplying the exponential suppression now reads: 
\be{s2}
s_2 =\left(\frac{\xi_{avg}}{c_{M}}\right)^{\Delta_\O - 1/2} e^{-\xi_{\mathcal{O}}\frac{\Delta_{avg}}{\xi_{avg}}}.
\ee
This is almost identical to the previous prefactor \eqref{s1}, except for a shift of the exponent of the first term: $\D_\O \to \D_\O -\frac{1}{2}.$ This extra factor of $-\frac{1}{2}$ comes about from the $\th$ integration performed in Sec~(3.2). The implications for this additional shift are not yet clear at this moment.

\bigskip

\section{Torus 2pt Functions and Bulk Quasinormal Modes} 

\subsection{BTZ QNM}
When a black hole is perturbed by a probe, the perturbation propagates with certain discrete frequencies due to fixed boundary conditions at the horizon and asymptotic infinity \cite{Birmingham:2001hc}. To study the black hole perturbation, one needs to solve $(\nabla^{2}-m^{2})\phi=0, $ in the black hole background. Here $m$ is the mass of the field $\phi$. The frequencies, called quasinormal modes, depend upon the parameters of the probe and the black hole.  Analytic computations for the decay rate of black hole perturbations were first carried out in the BTZ background in \cite{Birmingham:1997rj}. For a non-extremal BTZ, this computation yields: 
\be{quasi_mode_bulk}
 \omega_{L, R}=\pm\frac{k}{\ell}-2i\left(\frac{r_{+} \mp r_{-}}{\ell^{2}}\right)\left(n+\frac{1}{2}+\frac{1}{2}\sqrt{1+m^{2}\ell^2}\right).
 \ee
Here, $k$ is the angular momentum and $m$ the mass of the perturbation, $n$ takes integer values $0,1,2,\ldots$ and $r_\pm$ are the outer and inner horizons of the BTZ black hole. An interpretation of QNM in terms of dual thermal CFTs was first suggested in \cite{Horowitz:1999jd}, and the analysis for BTZ black hole perturbation and the decay of QNM in the context of 2D CFT was first carried out in \cite{Birmingham:2001pj}. The scalar probe of mass $m$ corresponds to an operator $\mathcal{O}$ with energy $E_{\mathcal{O}}$ in the 2d CFT, where
\be{Dm}
E_\O = h_\O + \bar{h}_\O = 1+ \sqrt{1+m^{2}\ell^2}.
\ee
Therefore, in AdS/CFT, the problem of black hole perturbation gets reduced to that of the perturbation of a thermal system by a probe and its eventual relaxation back to an equilibrium state. For a small perturbation, the process can be studied under the purview of linear response theory \cite{Birmingham:2002ph}. According to this analysis, the frequencies of the decay of perturbation, i.e., of the black hole quasinormal modes, are given by the poles of the correlation function of the perturbation operator $\mathcal{O}$ in momentum space. 

The relevant correlation function has the form
\begin{equation}
    \mathcal{D}(\omega, k)\propto \Big\vert\Gamma\left(h_{\mathcal{O}}+i\frac{p_{-}}{2\pi T_{R}}\right)\Gamma\left(\bar{h}_{\mathcal{O}}+i\frac{p_{+}}{2\pi T_{L}}\right)\Big\vert^{2}.
\end{equation}
Frequencies of the quasinormal modes are given by the poles of the gamma function and they are given by
\begin{eqnarray}\label{frq}
    \omega_{L}=k-4\pi iT_{L}(n+h_{\mathcal{O}}), \quad \omega_{R}=-k-4\pi iT_{R}(n+\bar{h}_{\mathcal{O}}),
\end{eqnarray}
where $T_{L/R}$ are related to $r_\pm$ as $T_{L,R} = 2 \pi \left( \frac{r_+ \pm r_-}{\ell^2} \right)$. Thus, the CFT answer \eqref{frq} matches exactly with the bulk analysis \eqref{quasi_mode_bulk}. 

\medskip

The off-diagonal structure constants can be obtained from the 2d CFT torus two-point function  \eqref{gamma_pole}. The poles of \eqref{gamma_pole} are given by 
\begin{equation}
    \omega_{n}=-i\left(\frac{12 E_{avg}}{c}-1\right)^{1/2}(n+E_{\mathcal{O}}). 
\end{equation}
For a non-rotating black hole, the Hawking temperature is given by 
\begin{equation}
    T_{H}=\frac{r_{+}}{2\pi \ell^{2}}.
\end{equation}
Relating the conformal weights $E$ with the mass of the black hole $M$ 
\begin{equation}
    M\ell=E-\frac{c}{12}=\frac{r_{+}^{2}}{8G\ell^{2}} \Rightarrow T_{H}=\frac{1}{2\pi \ell}\left(\frac{12 E}{c}-1\right)^{1/2}.
\end{equation}
Hence, the poles can be recast as,
\begin{equation}
\label{qnm_eth}
   \frac{\omega_{n}}{\ell}=-\frac{2\pi i}{\beta_{avg}}(n+E_{\mathcal{O}}), 
\end{equation} 
where $\beta_{avg}$ is the temperature corresponding to $E_{avg}$ and is identified with $T^{-1}_H$. From \eqref{frq}, setting $T_L=T_R=T_H$ for a non-rotating solution, one can readily see that 
\be{}
\w_n = \ell(\w_L+\w_R),
\ee
and there is no $k$ dependence as the probes are inserted at the same angular location and are only displaced in time. One thus sees that the off-diagonal three-point coefficients in a 2d CFT carry the imprints of BTZ QNM and provide a qualitative understanding of these QNM. 

\subsection{Singularity Structure of BMS Three-Point Coefficients}
In an effort to understand the analogs of QNM for the FSC solutions in flat space, we now focus on the singularity structure of the two different BMS off-diagonal structure constants that we have derived previously. Understanding QNM for FSCs has been obstructed up until now as it was not clear what boundary conditions should be used. The singularity structure of the off-diagonal structure constants now provides a way to understand QNM for FSCs. 

\medskip

In analogy to the BTZ case, we interpret the average three-point coefficients as denoting the transition of an FSC solution from an initial state with mass and angular momentum proportional to $\xi$ and $\Delta$ to a final state with $\xi^{\prime}$ and $\Delta^{\prime}$. We will consider the two cases (temporally and spatially separated) treated previously below. 

\paragraph{Temporally Separated Probes:}  For temporally separated probes, we assumed $\Delta_{i}=\Delta_{j}$. This choice also constrains our analysis to a case where the FSC transition takes place between two states with the same angular momentum. $\overline{\vert C_{\xi\mathcal{O}\xi^{\prime}}\vert^{2}}$ is thus a measure of the transition rate of an FSC state from $\vert\xi,\Delta\rangle$ to $\vert\xi^{\prime},\Delta\rangle$. During this transition, the field with energy $\zeta=\xi-\xi^{\prime}$ is emitted or absorbed (depending on the sign of $\zeta$). This process bears a close resemblance to the perturbation process of a black hole. The decay of the perturbation can be inferred from the poles of the transition rate. The black hole horizon has now been replaced with a cosmological horizon making the process even more interesting. The singularity structure of the off-diagonal three-point constant $\overline{\vert C_{\xi\mathcal{O}\xi^{\prime}}\vert^{2}}$, as can be seen from \eqref{eqD}, is given by a localized value of $\zeta$ that reads
\begin{equation}
    \zeta=i\xi_{\mathcal{O}}\sqrt{\frac{2\xi_{avg}}{c_{M}}}.
\end{equation} 
 In terms of FSC parameters, one has,
 \begin{equation}\label{sing}
     \zeta=i\xi_{\mathcal{O}}\sqrt{8GM_{avg}}=i\xi_{\mathcal{O}} {\hat{r}_+}.
 \end{equation} 
It is important to note that one doesn't get a discrete set of frequencies but just a single one. This collapse of the QNM in the flat space limit is expected from earlier work. If one thinks of this in terms of the flat space limit, the QNM of the BTZ black hole are frequencies of oscillations of a probe that is located in the region between $r=r+$ and $r\to\infty$. As one takes the flat space limit, the pole structure gets squeezed together as the outer radius gets pushed out. In the strict $\ell\to\infty$ limit, all these poles converge into a single pole as $r_+$ coincides with the boundary. The remnant of this pole structure is what we find in \eqref{sing}.

\paragraph{Spatially Separated Probes:} The expression $\overline{\vert C_{\Delta\mathcal{O}\Delta^{\prime}}\vert^{2}}$ denotes the emission of a particle with energy equivalent to $\lambda=\Delta-\Delta^{\prime}$. $\xi_{i}=\xi_{j}$ corresponds to a situation where the mass of the FSC before and after the transition remains the same. The probe thus carries away/in the angular momentum lost/gained in the process. The singularity structure in this case differs significantly from the earlier case. One can see that the frequencies of quasinormal modes are spread over a discrete spectrum, showing a similar behaviour as the analysis of BTZ quasinormal modes. As can be seen from the poles of the gamma function in the expression for the three-point structure constant \eqref{eqxi}, the singularity structure reads
  \begin{equation}
      \lambda=\pm i\sqrt{\frac{2\xi_{avg}}{c_{M}}}\left(n+\Delta_{\mathcal{O}}+\frac{\xi_{\mathcal{O}}\Delta_{avg}}{2\xi_{avg}}\right).
  \end{equation}
 Expressing this in terms of the FSC mass and angular momentum one obtains,
 \be{sing2}
     \lambda=\pm i\left[(n+\Delta_{\mathcal{O}})\sqrt{8GM}+\xi_{\mathcal{O}}J \sqrt{\frac{2G}{M}}\right] = \pm i\left[(n+\Delta_{\mathcal{O}})\hat{r}_+ +\xi_{\mathcal{O}}r_0 \right].
 \end{equation} 
One can interpret these as the spatial imprints of a perturbation of the cosmological horizon of the FSC when a probe $\O$ with BMS weights $(\D_\O,\xi_\O)$ is thrown at it. The flipping of the spatial and temporal directions from the point of the BTZ solution is evident from the fact that this answer mimics the usual BTZ QNM and arises from the poles of a gamma function, even though this is an answer arising from the calculation of a two-point function in an FSC background with purely spatially separated probes. The fact that we have the initial and final states of the FSC at the same mass and different angular momenta also indicates that it is the Virasoro sub-algebra of the BMS algebra which is important here and not the supertranslations. The spatially separated probes also indicate that the relevant underlying structure is the celestial circle at null infinity and hence the relevant two-point functions are determined via Diff(S$^1$).  

\bigskip

\paragraph{Recovering Answers in the Flat Space Limit.} Rather remarkably, as we will see below, one can recover the answers that we obtained using singularities from the off-diagonal BMS structure constants by invoking modular covariance of BMS torus two-point functions, from a limit on the QNM of the BTZ black hole. One can rewrite the left and right frequencies at zero angular separation of probes as
\be{eq:BTZQNMFRequencies}
\w^L_n = - 2i \left(\frac{r_++r_-}{\ell^2}\right)(n+h_L), \quad \w^R_n = - 2i \left(\frac{r_+-r_-}{\ell^2}\right)(n+h_R).
\ee
We now make the following substitutions as $\ell \to \infty$:
\be{}
h_L= \D_\O + \ell \xi_\O, \quad h_R = -\D_\O + \ell \xi_\O, \quad r_+\to \ell \hat{r}_+, \, r_-\to r_0.
\ee
Inserting this in \eqref{eq:BTZQNMFRequencies} one obtains
\bea{}
&& \w^L_n \to - i \xi_\O \hat{r}_+ - \frac{i}{\ell}\left[(n+\D_\O)\hat{r}_+ + r_0 \xi_\O \right] + \ldots \\
&& \w^R_n \to - i \xi_\O \hat{r}_+ + \frac{i}{\ell}\left[(-n+\D_\O)\hat{r}_+ + r_0 \xi_\O \right] + \ldots
\eea
and similarly for $\w_R$. In the limit $\ell\to\infty$, one obtains
\bea{}
&& \lim_{\ell\to\infty} \w_n^{L,R} = - i \xi_\O \hat{r}_+ \pm \frac{i}{\ell}\left[(\pm n+\D_\O)\hat{r}_+ + r_0 \xi_\O \right], \\
&& \Rightarrow \zeta_n = \w_n^{L} + \w_{-n}^{R}, \quad \lambda_n = \ell  (\w_n^{L} - \w_{-n}^{R}),
\eea
where we have used our previous results \eqref{sing} and \eqref{sing2}. Thus, the leading and next-to-leading pieces of the BTZ QNM in the flat limit reproduce the pole structure we obtained from the BMS off-diagonal three-point coefficients. It is interesting to see that these are encoded in two different off-diagonal structure constants, one for temporally separated probes and one for spatially separated probes. It is perhaps expected that the flat limit of the BTZ QNM would lead us to the answer of the collapse of the QNM in the flat limit and the answer $\overline{\vert C_{\xi\mathcal{O}\xi^{\prime}}\vert^{2}}$ recovered from the temporally separated probes. But that the subleading piece of the BTZ QNM in the flat limit, with the non-trivial pole structure, is contained in $\overline{\vert C_{\Delta\mathcal{O}\Delta^{\prime}}\vert^{2}}$ is surprising.

\bigskip
 
\section{Concluding Remarks}

\subsection{Summary}
In this work, we have investigated modular properties of 2d BMSFTs. Our goal was to exploit the covariance of the BMS torus two-point function to arrive at expressions for off-diagonal structure constants of the 2d BMSFT. Like in usual 2d CFTs, the knowledge of all structure constants of a theory, along with the spectrum of primaries and the central charges, is enough to completely characterize the theory using bootstrap methods outlined, e.g., in \cite{Bagchi:2016geg,Bagchi:2017cpu}. We have arrived successfully at (averaged expressions of) two different classes of structure constants  $\overline{\vert C_{\xi\mathcal{O}\xi^{\prime}}\vert^{2}}$ and $\overline{\vert C_{\Delta\mathcal{O}\Delta^{\prime}}\vert^{2}}$. We used the two-point functions of temporally and spatially separated probes, respectively, and also assumed equal background FSC angular momenta in the first case and equal FSC mass in the second case after a perturbation. These two expressions constitute the main results of our work and are given in \eqref{eqD} and \eqref{eqxi}. 

\medskip

There is rich physics hidden in the expressions of these off-diagonal structure constants. We saw that in both cases the off-diagonal structure constants were exponentially suppressed and this indicates that these BMSFTs follow a notion of the eigenstate thermalization hypothesis. The exponential suppression, however, had interesting additional pieces as compared to 2d CFTs, which led us to conclude that although the systems would thermalize, this process of thermalization would possibly be slower compared to their 2d CFT counterparts. The fact that the spatially separated probes also led to off-diagonal structure constants with exponential suppression was reminiscent of the spatial version of chaos seen earlier in 2d BMSFTs in \cite{Bagchi:2021qfe}. Finally, the singularity structure of these expressions \eqref{eqD} and \eqref{eqxi} were examined and connections were made to QNM of FSC solutions in the asymptotically flat bulk theory. We recovered the answers of the singularities through a flat limit of the BTZ QNM. Very surprisingly, we were able to match both the leading and subleading pieces of the large $\ell$ answer to the singularities of the two different classes of structure constants. 

\subsection{Discussions}
In our pursuit of a holographic dual for asymptotically flat spacetimes, we have investigated theories where the speed of light goes to zero. These Carrollian field theories and specifically Carrollian CFTs (see e.g \cite{Bagchi:2019xfx, Bagchi:2019clu, deBoer:2021jej}) provide a challenge to standard rules of relativistic QFTs as lightcones collapse and notions of causality and unitarity are challenged. These apparently bizarre theories have now also come up in many real-life scenarios, including the theory of flat bands \cite{Bagchi:2022eui} and fractons \cite{Bidussi:2021nmp} in condensed matter physics, cosmology \cite{deBoer:2021jej},  as well as in fluids moving with very high velocities e.g. in modeling of the quark-gluon-plasma \cite{Bagchi:2023ysc}. Understanding aspects of thermalization of these theories is thus an important problem and one where we have made some headway in our present paper. It does seem that structure constants of the 2d Carroll CFTs obey ETH and thus a Carroll CFT should reach thermal equilibrium, although with some interesting changes from that of a usual 2d CFT. This indicates that the possible issues about the non-existence of an equilibrium state for Carrollian field theories are perhaps not as serious as initially anticipated. Of course, our analysis is only for $d=2$, and further investigation is needed. However, our analysis in this work is a promising start for these future investigations. 

\medskip
Our other major contribution in this work is a proposal for the analogs of quasinormal modes for the FSC solutions of asymptotically flat spacetimes. This has been a long-standing problem because of the issue of boundary conditions. Adding perturbations to an FSC with e.g. a bulk scalar field, it has been very unclear what boundary conditions to impose on the cosmological horizon and most obvious attempts have not been successful. We have circumvented this issue by looking at the dual field theory and the poles of the thermal two-point functions which were embedded in our formulae for the off-diagonal structure constants. While this indirect method yielded surprisingly nice answers which matched with the flat limit of the BTZ QNM, it would be good to have further cross-checks of the spectrum we have uncovered. With the help of the expected QNM spectrum, one may be able to reconstruct what the ``proper" boundary conditions for the probe bulk field would be. Other methods, like using the Selberg zeta-function \cite{Keeler:2018lza}, would also be useful \cite{wip}. 

\medskip

The matching of the leading term of the $\ell\to\infty$ limit of the BTZ QNM with the singularity structure from the temporally separated probe in the 2d BMSFT is something that was not unexpected as we have stressed before. But the matching of the subleading term in the $\ell\to\infty$ limit with the singularities of the spatially separated probe seems surprising. The off-diagonal structure constants are very different in terms of the field theory. They denote temporal and spatially separated probes. They also carry information on transitions between FSCs of equal angular momentum and unequal mass in the first case and equal mass and unequal angular momentum in the second case. Why these two apparently very different three-point coefficients from different parts of the BMSFT phase space should carry information arising from the same BTZ QNM is a puzzle that needs further clarification. This looks similar to the Carroll expansion of the relativistic field theory (say scalar theory) action in a series where the speed of light $c$ controls the expansion parameter. There the leading term is the so-called electric term and the sub-leading one (with certain modifications) is the magnetic one. At first sight, these electric and magnetic theories don't seem related to each other at all, but they arise in the limit of the same relativistic theory. Perhaps there is something similar at play here as well. We hope to report on this and other aspects of perturbations of the FSC background and its formulation in terms of the dual 2d BMSFT in the near future.

\bigskip \bigskip

\subsection*{Acknowledgements}
We thank Diptarka Das, Daniel Grumiller, Cynthia Keeler, Victoria Martin, Rahul Poddar for interesting discussions and comments. 

\smallskip

During the course of this work, AB was partially supported by a Swarnajayanti fellowship (SB/SJF/2019-20/08) from the Science and Engineering Research Board (SERB) India, the SERB grant (CRG/2020/002035), a visiting professorship at \'{E}cole Polytechnique Paris, a distinguished visiting professorship at NORDITA Stockholm, and a Royal Society of London international exchange grant with the University of Edinburgh. AB also acknowledges the warm hospitality of the Niels Bohr Institute, Copenhagen, the University of Edinburgh, UK, ULB Brussels and NORDITA, Stockholm, during various stages of this work. SM is supported by grant number 09/092(1039)/2019-EMR-I from Council of Scientific and Industrial Research (CSIR). The research of MR is supported by the Austrian Science Fund (FWF) project P32581. MR also acknowledges the warm hospitality of Harvard University, the Okinawa Institute of Science and Technology (OIST), the Perimeter Institute for Theoretical Physics (PI), and the Yukawa Institute for Theoretical Physics (YITP) during various stages of this work.

\newpage
\bibliographystyle{JHEP}
\bibliography{bibli}	
\end{document}